\documentclass[prb,twocolumn,showpacs,preprintnumbers,amsmath,amssymb,a4paper,superscriptaddress]{revtex4}

\usepackage{graphicx}
\usepackage{dcolumn}
\usepackage{bm}

\begin{document}


\title{Magnetoconductance switching in an array of oval quantum dots}

\author{C. Morfonios}
\affiliation{Theoretische Chemie, Institut f\"{u}r Physikalische Chemie, Universit\"{a}t Heidelberg, Im Neuenheimer Feld 229, 69120 Heidelberg, Germany}
\author{D. Buchholz}
\affiliation{Theoretische Chemie, Institut f\"{u}r Physikalische Chemie, Universit\"{a}t Heidelberg, Im Neuenheimer Feld 229, 69120 Heidelberg, Germany}
\author{P. Schmelcher}
\affiliation{Theoretische Chemie, Institut f\"{u}r Physikalische Chemie, Universit\"{a}t
Heidelberg, Im Neuenheimer Feld 229, 69120 Heidelberg, Germany}
\affiliation{Physikalisches Institut, Universit\"{a}t Heidelberg, Philosophenweg 12, 69120
Heidelberg, Germany}


\date{\today}

\begin{abstract}
Employing oval shaped quantum billiards connected by quantum wires as the building blocks of a linear quantum dot array,
we calculate the ballistic magnetoconductance in the linear response regime.
Optimizing the geometry of the billiards, we aim at a maximal finite- over zero-field ratio of the magnetoconductance.
This switching effect arises from a relative phase change of scattering states in the oval quantum dot through the applied magnetic field, 
which lifts a suppression of the transmission characteristic for a certain range of geometry parameters.
It is shown that a sustainable switching ratio is reached for a very low field strength,
which is multiplied by connecting only a second dot to the single one.
The impact of disorder is addressed in the form of remote impurity scattering,
which poses a temperature dependent lower bound for the switching ratio,
showing that this effect should be readily observable in experiments.
\end{abstract}

\maketitle


\section{Introduction}
The ability to reduce the size of electronic circuits to the nanometric scale 
has lead to increasing interest in the properties of electron transport in the mesoscopic regime,
and its dependence on externally tuned parameters.
Formation of two-dimensional (2D) structures of controllable geometry at semiconductor interfaces, so called electron billiards, 
sets the experimental grounds for investigating phase coherent transport of electrons.
Open semiconductor quantum billiards serve as artificial scatterers of highly tunable characteristics and have pioneered the understanding
of the underlying physics, on both experimental and theoretical grounds.
They are used to demonstrate and investigate a series of interesting phenomena on the mesoscopic level, such as shot noise in transport through charged dots \cite{Blanter2000,Golubev2004}, Fano resonances \cite{Clerk2001,Fang2008,Zeng2002,Mendoza2008}, Andreev tunneling and reflection \cite{Schmelcher2005,Fazio1998,Recher2001}, decoherence in ballistic nanostructures \cite{Knezevic2008,Bird2003} as well as classical to quantum transitions and imprints of nonlinear dynamics \cite{Nazmitdinov2002,Richter2002,Schmelcher2005,Liu2006}.
Further, the geometry of a conducting structure is shown to have a major impact on the resulting transport phenomena \cite{Buchholz2008,Weymann2008}.
The magnetoconductance of such nanodevices proves as an essential signature for the underlying interference phenomena
and has therefore been studied extensively \cite{Fang1968,Buttiker1986,Marcus1992,Baranger1993}.
The Aharonov-Bohm effect \cite{Aharonov1959} is directly observed in systems of quantum rings \cite{Buttiker1986,Kalman2008,Jana2008,Zelyak2008}, 
but also plays a central role in describing magnetoconductance fluctuations in more complex mesoscopic systems in weak magnetic fields \cite{Drouvelis2007,Wang1994}.
At higher magnetic field strengths the quantum Hall effect sets in, accounting for a steplike varying magnetoconductance, formation of edge states and characteristic multi-channel fluctuations in the transmission spectra \cite{Rotter2003,Kawarabayashi2008,Johnson1992}.
Localization effects and conductance fluctuations manifest themselves in a large variety of open quantum dot systems, 
regardless of whether ballistic \cite{Marcus1992,Baranger1993,Jalabert1990,Nazmitdinov2002,Brezinova2008,Brouwer2008} or diffusive \cite{Hastings1994,Golubev2006a,Elhassan2001,Brouwer2008} transport is considered.
Assembling individual dots into coupled arrays or lattices gives rise to new features of the system's overall response, depending on the type and strength of coupling \cite{Drouvelis2007,Lobos2006,Cai2007,Teng2007,Wegewijs1999,Asai2005}.
Of particular interest are systems where the interplay between the various effects of electron transport mentioned above can be used to achieve a tunable quantum conductance,
in terms of designing the size, shape and material specific features of the conducting device, 
as well as varying macroscopically accessible parameters such as externally applied fields, temperature, and gate voltages controlling the coupling strength between constituents \cite{Drouvelis2007,Buchholz2008,Waugh1995,Asai2005,Elhassan2001,Elhassan2004,Bird2003}.

In this article we exploit the dependence of the conductance on the geometry of a 2D electron billiard and
examine its functionality as a switch when a magnetic field is turned on.
Employing oval shaped billiards as the building blocks of a linear quantum dot array, 
we aim at a maximal finite- over zero-field ratio of the conductance by optimizing the system within an achievable parameter range.
The switching effect arises from the phase changing effect of the applied field, 
which raises a suppression of transmission present for a certain deformation of the oval.
The assembly of dots into a chain eventually leads to banded transmission spectra for a large number of dots, 
with details depending on the interdot lead length.
The conductance, taken as the thermally averaged transmission function, oscillates with increasing field strength; 
at higher fields edge states form, which conduct ideally.
The switching ratio corresponding to the first magnetoconductance maximum acquires a multiple value by adding one more oval to the single one, 
while it fluctuates for further added dots.
The impact of impurities may enhance or weaken the switching effect, whether or not they block the leads coupled to the dots,
imposing a temperature dependent lower bound on the switching ratio in the presence of weak disorder.

In section II the setup and geometry of the 2D structure are specified and the theoretical framework as well as computational approach are presented.
In section III the main features of the obtained transmission spectra are discussed, 
along with a description of the underlying mechanisms. 
This is followed by an analysis of the switching ratio in dependence of the deformation of the billiard shape, 
the magnetic field strength and the length of the multidot chain at different temperatures, 
in order to determine a device setup optimal for switching, within an achievable parameter range.
Finally, the modification of the switching ratio in the presence of disorder is studied.
Section IV provides a summary of results, 
concluding on the functionality of the switching mechanism.

\section{Setup and computational approach \label{setup}}

The confining potential of the single dot is assumed to be of hard wall character, leading to Dirichlet boundary conditions for the wave function. 
We use an oval billiard, whose shape is parametrized as \cite{Berry1981}
\begin{eqnarray}
x(\phi) & = & R \left [\left (\frac{\delta}{2} + 1\right) \sin(\phi) + \frac{\delta}{6} \sin(3\phi)\right] \nonumber \\ 
y(\phi) & = & R \left [\left (\frac{\delta}{2} - 1\right) \cos(\phi) - \frac{\delta}{6} \cos(3\phi)\right]
\end{eqnarray}
where  $\phi \in [0,2\pi]$. 
The parameter $\delta$ tunes the deformation of the dot, which becomes a circular billiard of radius $R$ if $\delta = 0$.
In this case the classical dynamics of the closed system is integrable, whereas for $\delta > 0$ it becomes non-integrable with mixed phase space \cite{Berry1981,Makino1998}.
For reference with respect to the device specific parameters, a mesoscopic size of $R = 220~\rm{nm}$ is employed.
At the right and left ends of the elongated structure semi-infinite leads of width $W = 0.3~R$ are connected, 
representing the coupling to electron reservoirs.
The use of semi-infinite leads models the ideal case of vanishing reflection of the electrons upon reaching the reservoirs.
In the multidot case this single cavity is replaced by a chain of $N$ identical oval dots connected to each other through leads of length $L$, where $L$ equals the distance between adjacent oval edges,
aligned with the semi-infinite leads possessing the same width.
\begin{figure}
    \begin{center}  
      \includegraphics[width=0.45\textwidth]{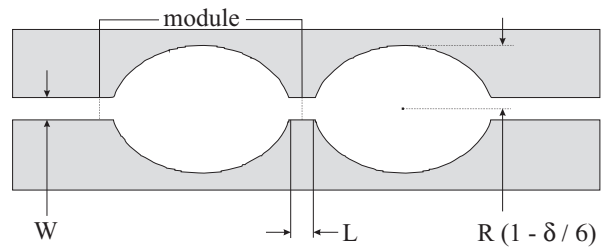}
     \end{center}
  \caption{Geometry of the billiard for $N = 2$, $\delta = 0.5$, $L = W = 0.3~R$, 
           consisting of the repeated module and the outer leads which represent the connection to reservoirs (see text).}
  \label{fig1}
\end{figure}
Fig.~\ref{fig1} provides a picture of the 2D structure for $N = 2$ connected dots.

Restricting ourselves to low temperatures and a small system size we neglect inelastic processes, 
and do not account for electron-electron or electron-phonon interactions.
The single particle Hamiltonian is, within an effective mass approach, of the form
\begin{eqnarray}
H = \frac{({\bf{p}} - e {\bf{A}})^2}{2 m_{\mathrm{eff}}} + V(\bf{r}) ~~,
\end{eqnarray}
where we choose a value of $m_{\mathrm{eff}} = 0.069~ m_e$ corresponding to GaAs, $m_e$ denoting the electron mass.
$V(\bf{r})$ is the hard wall potential, and the vector potential $\bf{A}$ produces a magnetic field perpendicular to the plane of the structure,
over which it is homogeneously extended with strength $B$, dropping off linearly to zero in the exterior leads.
We will concentrate on the magnetoconductance switching effect at a very low magnetic field strength ($\sim 0.02~\rm{T}$), 
where the Zeeman splitting for GaAs ($\sim 3.6~\rm{\mu eV}$) is negligible ($\sim 0.1\%$) with respect to the Fermi energies we consider, 
and therefore do not take into account the coupling of the electronic spin to the magnetic field.
The Hamiltonian is discretized on a tight-binding lattice, with the magnetic vector potential incorporated through Peierls' substitution.
The coupling of the system to the external semi-infinite leads placed on the left ($l$) and right ($r$) of the billiard is described by self-energies ${\bf \Sigma}_{l/r}$, which are analytically obtained for $B = 0$ and contribute non-Hermitian blocks to the Hamiltonian matrix.
From the single-particle Green's function of the system
\begin{eqnarray}
{\bf G}(E) =[ E{\bf I} - ( {\bf H} + {\bf \Sigma}_r  + {\bf \Sigma}_l ) ]^{-1}
\end{eqnarray}
the part ${\bf G}_{rl}$ describing the propagation from the left to the right lead 
is computed using a parallel implementation of the recursive Green's function method (RGM),
where a decomposition scheme among communicating processors allow for the computation to be done in a parallel manner \cite{Drouvelis2006}.
In the multidot case the chain is built up by a repeated module, 
which consists of the oval cavity with leadstubs of length $L/2$ on the right and left (see Fig.~\ref{fig1}).
Having found ${\bf G}_{rl}$ for one module,
we calculate the Green's function connecting the two outer leads using a modular variant of the RGM, 
which was originally presented in Ref.~\onlinecite{Rotter2000}.
In this algorithm the Green's function of the joined module is calculated using the Dyson equation.
The transmission of the device is finally evaluated via the Fisher-Lee relations \cite{Fisher1981},
$T(E) = Tr \left({\bf  \Gamma}_r {\bf G} {\bf \Gamma}_l {\bf G}^\dagger \right)$,
with ${\bf \Gamma}_{l/r} = i ( {\bf \Sigma}_{l/r} - {\bf \Sigma}^\dagger_{l/r})$.
It is worthwhile noting that in the two-terminal device we encounter, even in the presence of a magnetic field, 
the transmission function is symmetric under the exchange of the contact leads, 
i.e. the transmission from left to right equals that from right to left \cite{Buttiker1988}.
The computed Green's function of the system is also used to calculate the local density of states (DoS) at site $\bf{r}$ through the relation
$\rho({\bf{r}},E)=\langle{\bf{r}}\vert{\bf{A}}(E)\vert{\bf{r}}\rangle/2\pi$, 
where $\bf{A}={\bf{G}}{\bf{\Gamma}}{\bf{G}}^\dagger$ is the spectral function 
and ${\bf{\Gamma}}$ generally a weighted sum of ${\bf \Gamma}_{l}$ and ${\bf \Gamma}_{r}$ 
according to the Fermi distributions of incoming states in the two leads. 
In the cases presented here, we have chosen ${\bf{\Gamma}} = {\bf \Gamma}_{l}$,
i.e. $\rho({\bf r},E)$ corresponds to the probability density resulting from an incoming monochromatic wave of energy $E$ from the left lead.

The calculated transmission determines the macroscopically measurable conductance of the device.
In the linear response regime at low temperature $\Theta$ the conductance for given Fermi energy $E_F$ can be obtained by 
the Landauer formula \cite{Landauer1970,Buettiker1984}:
\begin{eqnarray}
G(E_F) = \frac{2e^2}{h} \int_{-\infty}^{+\infty} \! T(E)F_{th}(E_F;E) \, dE
\end{eqnarray}
with
\begin{eqnarray}
F_{th}(E_F;E) & \equiv & -\frac{\partial f(E_F;E)}{\partial E} 
             \cr & = & \frac{1}{4k_B\Theta}sech^2\left(\frac{E - E_F}{2k_B\Theta}\right)
\end{eqnarray}
where $f(E_F;E)$ is the Fermi distribution function centred around $E_F$,
and thus it essentially equals the thermally averaged transmission around the electron Fermi energy, 
with a width determined by the temperature $\Theta$.

\section{Results\label{results}}

\subsection{Transmission spectra \label{spectra}}
Concentrating on the transmission in the deep quantum regime, 
we restrict the energy of the incoming electrons 
such that only the transversal ground state of the leads is energetically available.
Thus the dimensionless channel number $\kappa = kW/\pi$, where $k = \sqrt{2m_{\mathrm{eff}} E }/\hbar$,
takes on values in the range $1 < \kappa < 2$, that is, in the first channel of transmission.
For the size of the device specified, this corresponds to a Fermi energy in the range $1.2~{\rm{meV}} < E_F < 5~{\rm{meV}}$.
A detailed analysis of the transmission within the first channel in terms of the quantum states in the single oval dot 
has been presented in Ref.~\onlinecite{Buchholz2008}.
Here we focus on the modification of the transmission when dots are connected to form an array,
as well as the conductance of the device (where the details of the transmission are thermally averaged out)
as a function of the geometry parameters and the magnetic field strength.
\begin{figure*}
    \begin{center}  
      \includegraphics[width=0.98\textwidth]{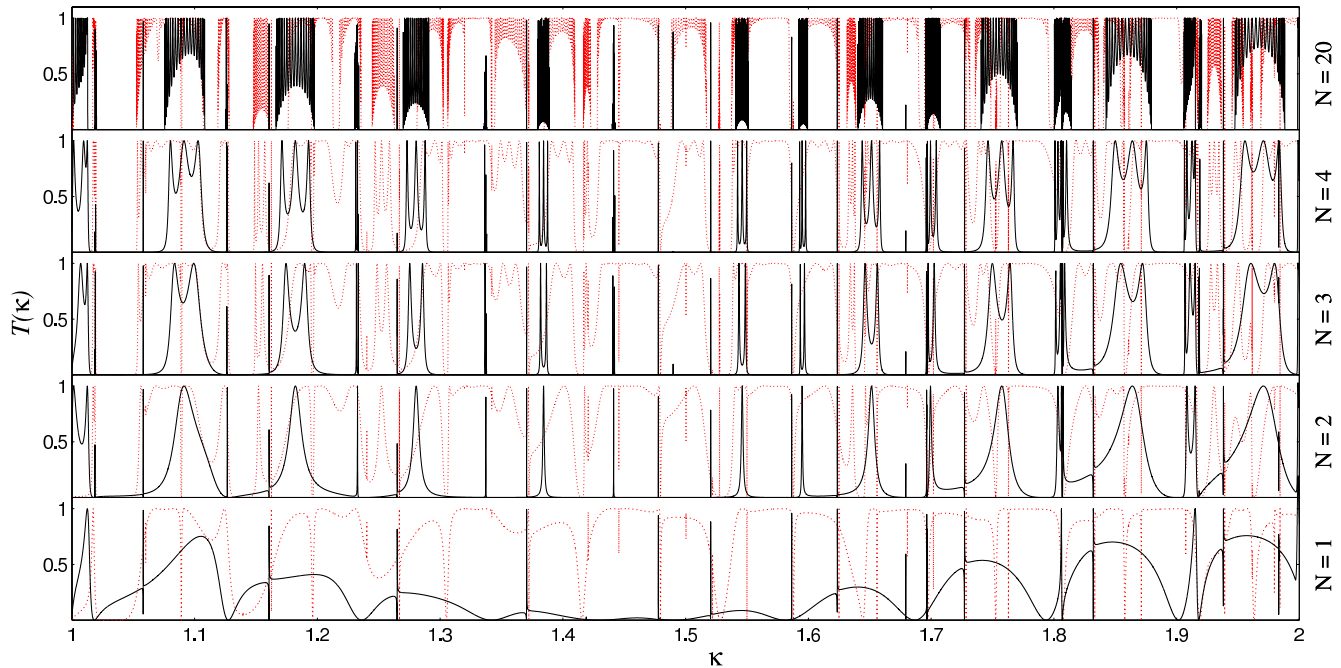} 
     \end{center}
  \caption{(Color online) Transmission spectra in the first transversal channel for varying number of dots $N$ with deformation parameter 
           $\delta = 0.5$ and connecting lead length $L = W = 0.3~R$, at $B = 0$ (solid black line) and $B = B_c \approx 20~mT$ (dotted red line).}
  \label{fig2}
\end{figure*}
The zero- and finite-field transmission $T(\kappa)$ through the device is shown in Fig.~\ref{fig2} 
for different numbers of dots $N$ in the chain,
with deformation parameter $\delta = 0.5$ and interdot distance $L = W$.
As the channel number $\kappa$ measures the wave number in units of $\pi/W$, 
$T(\kappa)$ depends only on the ratio $W/R$.
Our calculations show that changing $W/R$ within $0.2 \lesssim W/R \lesssim 0.4$ introduces mainly a shift in $T(\kappa)$
according to the implicit energy scaling,
i.e. the transmission is largely determined by the geometry of the billiard and not by the leadwidth.
For values of $W/R > 2$, the transmission obviously has to acquire the value of the unperturbed quantum wire.
In the following we restrict ourselves to the case of $W/R = 0.3$.
The zero-field transmission in the single dot case consists of a rather smoothly varying background, 
on which sharp Fano resonances are superimposed.
In the multi-oval case these sharp resonances are $N$-fold split 
(this very small splitting is generally not resolved on the scale of Fig.~\ref{fig2}), 
while, additionally, Breit-Wigner (BW) type resonances of varying width emerge,
firstly for $N = 2$, and subsequently undergo a splitting into $N-1$ sub-peaks for an array of $N$ dots.
For sufficiently many dots (represented in Fig.~\ref{fig2} by the case of $N = 20$), 
the multiply split resonances saturate into bands of densely positioned peaks,
which is reminiscent of the band structure of energy levels in a periodic quantum system.
In the presence of the weak field the smooth background transmission is overall increased, 
the sharp resonances are slightly shifted in energy
and the transmittive bands for large $N$ are broader.

The sharp Fano resonances of the single dot case 
correspond to quasibound states that are strongly localized within the dot, 
at energies which coincide with eigenenergies of the closed oval billiard without attached wires.
There is a series of equidistant resonances corresponding to the different longitudinal modes for each given excited transversal mode in the oval \cite{Buchholz2008}.
The spatial confinement of such states decouples them from the leads 
and consequently they do not contribute significantly to the conductance of the device.
The states that can provide a substantial contribution to the conductance
are those with a longitudinal spatial extension unto the openings of the leads.
They are strongly coupled to the leads of the open system in the case of constructive interference at the openings,
leading to a broad transmission maximum.
We refer to these states, which extend from the oval into the leads, as leaking states.
The number of leaking states is determined by the allowed excitations inside the cavity 
subject to the constraint of the energy being within the first channel.
The interference of leaking states belonging to different transversal modes 
generates broad humps in the single dot transmission, where the transmission is substantial over a finite energy interval (constructive interference), 
separated by points of vanishing transmission (destructive interference). 
The slowly varying envelope behavior of the transmission spectrum exhibits a wide energy range where the overall transmission is strongly suppressed.
For the specific shape of the cavity corresponding to the chosen value of $\delta = 0.5$, 
this suppression valley is centered around the middle of the first channel.

\begin{figure}
    \begin{center}  
      \includegraphics*[width=0.48\textwidth]{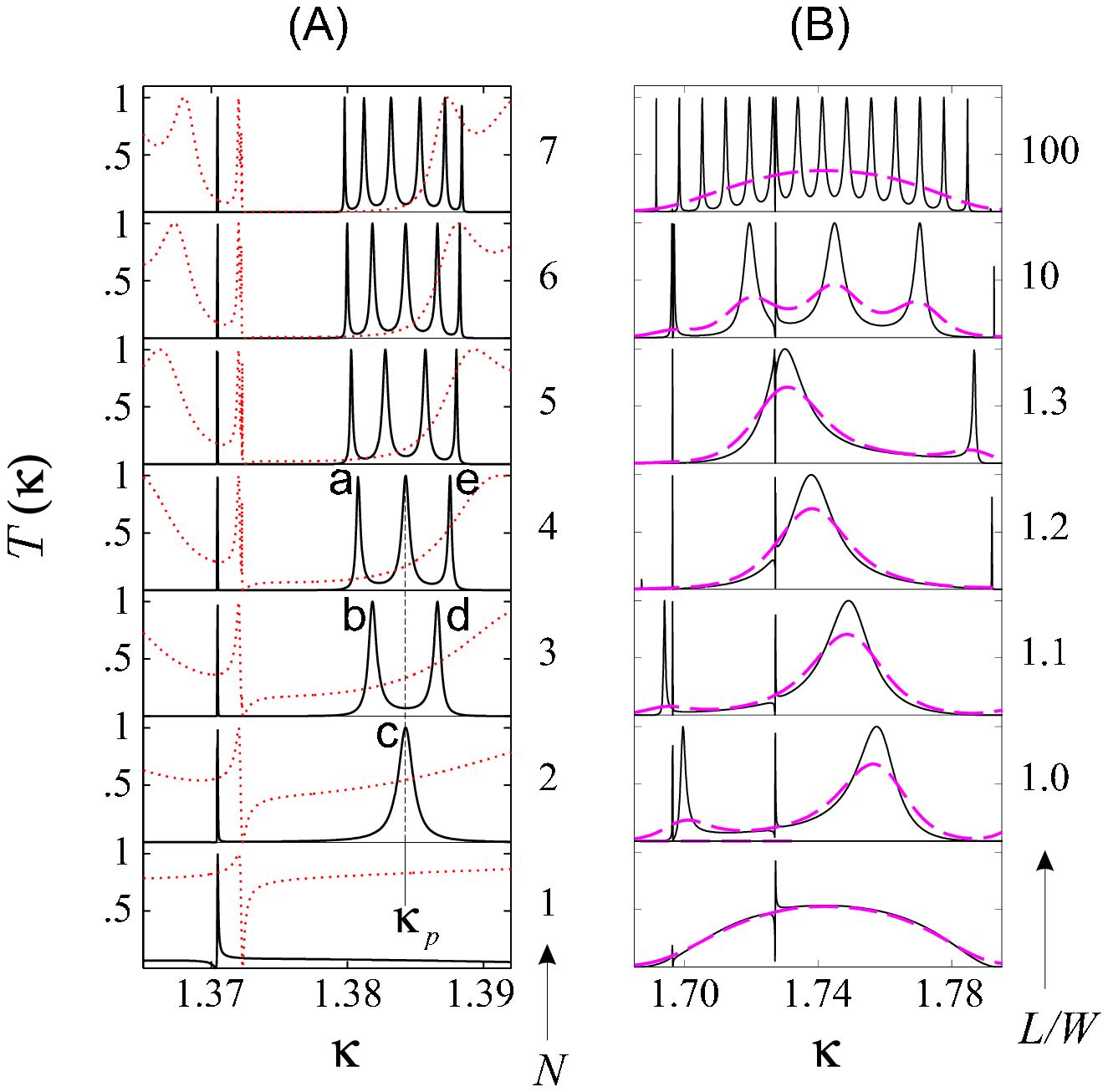} 
     \end{center}
   \caption{(Color online) (A): $T(\kappa)$ for $B = 0$ (solid black line) and $B = B_c$ (dotted red line) 
            for $L = W$ and varying $N$, in the vicinity of $\kappa_p \approx 1.384$, 
            with labels a, b, c, d, e for the resonances referred to in the text,
            (B): $T(\kappa)$ (solid black line) and $g(\kappa~;~\Theta = 0.2~\rm{K})$ (dashed magenta line)
            for a single dot (bottom) and for two dots with varying bridge length $L$,
            within a small window of the channel number $\kappa$ covering the energy range of a single dot smooth hump.}
   \label{fig3}
\end{figure}
In order to analyze the transmission of the multidot chain, 
in Fig.~\ref{fig3}~(A) we focus on the transmission around the BW resonance appearing for $N = 2$ 
at $\kappa = \kappa_p \approx 1.384$, and show its $(N-1)$-fold splitting for increasing $N$.
Also the sharp Fano resonance just below is included, whose splitting (of the order of $\Delta\kappa \sim 2\times10^{-5}$ or $\Delta E \sim 0.1~\rm{\mu eV}$) remains unresolved even at this scale.
The $N$-fold splitting of the Fano resonances 
is a consequence of the degeneracy of the confined single dot eigenstates in the case of $N$ dots, 
which are coupled very weakly through the connecting lead due to their strong localization within the ovals.
It is thus similar to the splitting of the energy levels of atoms brought together to form a weakly bound molecule,
with an energy split proportional to the interatomic coupling \cite{Neumann1929}.
The BW type resonances of the multidot case, 
which are narrower (wider) at energies where the single dot transmission $T^{(N=1)}(\kappa)$ is lower (higher),
are of different origin:
They arise from the resonant tunneling of the incoming wave through the system of the ovals and the connecting bridges.
Indeed, the emergence of these resonances and their ($N-1$)-fold splitting 
can effectively be deduced from the 1D scattering through $N$ potential barriers 
(or equivalently, $N-1$ resonators),
where the transmission amplitude of scattering through each barrier possesses an energy dependent norm and phase.
Two barriers $\alpha$, $\beta$ with transmissions $T_\alpha$, $T_\beta$ give the total transmission
\begin{equation}
T_{\alpha\beta} = T_{\alpha}T_{\beta} / [1 + R_{\alpha}R_{\beta} - 2\sqrt{R_{\alpha}R_{\beta}}cos\theta]~~ ,
\label{eq_combined_scatterers}
\end{equation}
where $R_{\alpha/\beta} = 1 - T_{\alpha/\beta}$ and $\theta$ is the phase shift acquired by reflection from $\beta$ to $\alpha$ and back to $\beta$. For $T_{\alpha} = T_{\beta} = T^{(N = 1)}$ and $\theta \varpropto \kappa$ this gives rise to resonance peaks in $T^{(N = 2)}$ which are equidistant in $\kappa$
and have a width that increases with $T^{(1)}$.
In our case though, due to the structure of the ovals that constitute the barriers, 
the phase shift $\theta$ is not linear in $\kappa$.
This perturbs the periodicity of the resonances, as we observe for $T^{(2)}(\kappa)$ in Fig.~\ref{fig2},
or equivalently, yields an energy dependent effective resonator length $\tilde{L}(\kappa) \varpropto \theta(\kappa)/\kappa$.
Formula (\ref{eq_combined_scatterers}) can be iterated to obtain the transmission for 
$N \geqslant 2$ ovals, i.e. 
$T^{(N)} = T_{\alpha\beta}(~T_{\alpha}=T^{(1)}, T_{\beta}=T^{(N-1)}; \theta_{\alpha,\beta}=\theta_{1,N-1}~)$,
where $\theta$ now results from reflections between $1$ and $N-1$ barriers.
The $(N-1)$-fold splitting of the $T^{(2)}$ resonance, 
shown in Fig.~\ref{fig3}~(A), 
and the saturation into a band in the transmission spectrum for large $N$, 
are then reproduced for a system that is symmetric under the exchange $\alpha \leftrightarrow \beta$ 
(which, in our case, renders the dots identical),
provided that the phase difference between transmission and reflection amplitude of the single barrier is equal to $\pm \pi/2$, as is the case for the single oval with symmetric leads.
Varying the resonator length modifies the conditions for resonant transmission by shifting the resonances in energy
and changing their periodicity.
In Fig.~\ref{fig3}~(B) the transmission through $N = 2$ connected dots,
as well as the normalized conductance at $\Theta = 0.2~\rm{K}$, 
are plotted over the energy range of a single dot transmission hump,
for varying connecting bridge length $L$.
With a slight increase in $L$ ($L/W = 1.0, 1.1, 1.2, 1.3$) the BW resonances are shifted to lower energy,
and for longer bridges ($L/W = 20, 100$) the number of resonances in the same interval increases.
We notice that the center positions of the (split) Fano resonances are unaffected by the variation of the bridge length.
Detailed features of the transmission lineshape, such as the Fano resonances and the BW resonance peaks for large $L$, 
are washed out by thermal averaging, making their contribution to the conductance negligible compared to the smooth background.

As we see in Fig.~\ref{fig3}~(A), the addition of a dot to the existing chain at resonance energy, 
lowers the transmission from unity to the single oval value $T^{(1)}(\kappa)$ at that energy 
(note that the transmission at the dips between the resonances can acquire values even lower than $T^{(1)}(\kappa)$~).
In particular, the transmission at the energy position of the central resonance at $\kappa = \kappa_p$ 
oscillates between unity and $T^{(1)}(\kappa_p)$ with even and odd $N$, respectively: 
$T^{(N\rm{\textit{even}})}(\kappa_p) = 1$, $T^{(N\rm{\textit{odd}})}(\kappa_p) = T^{(1)}(\kappa_p)$.
Furthermore, the resonances for each $N$ are positioned symmetrically around $\kappa_p$, 
so that the forming bands in the transmission for large $N$ are centered around the $T^{(2)}(\kappa)$ resonance peaks.
\begin{figure*}
  \begin{center}  
      \includegraphics[width=0.98\textwidth]{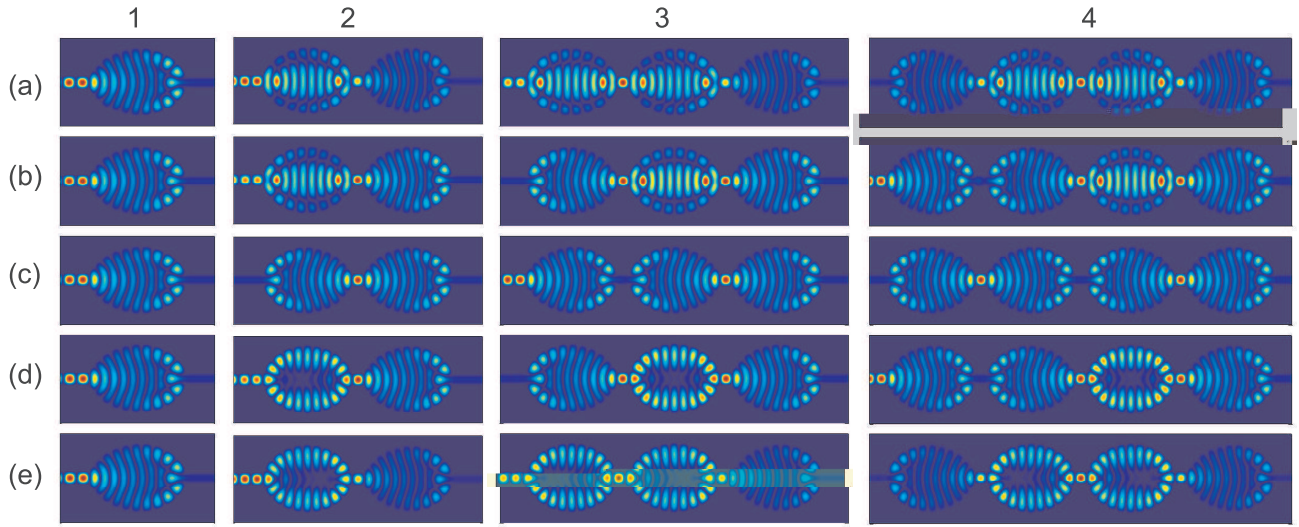}
     \end{center}
  \caption{(Color online) Zero-field local DoS for $N = 1,2,3,4$ dots with $\delta = 0.5$ and $L = W$ 
          for energies in the vicinity of the $(N-1)$-fold split resonance peak at 
          $\kappa = \kappa_p \approx 1.384$, with incoming electron on the left. 
          Rows (a), (b), (c), (d), (e) correspond to the energies of the resonances labeled with the same letters in 
          Fig.~\ref{fig3}~(A).
          The colormap for the DoS in each sub-plot is normalized to its maximal value, 
          further the color maps the square root of the DoS to enhance contrast.}
  \label{fig4}
\end{figure*}
In Fig.~\ref{fig4} this behavior of the transmission function for varying number of dots is illustrated in terms of the states forming in the system for $N = 1,2,3,4$ dots, 
by plotting the zero-field local DoS at the energies (rows a,b,c,d,e) of the resonance peaks labeled (with corresponding letters) in Fig.~\ref{fig3}~(A), for an electron incident on the left.
The spatial oscillations of the DoS in the incoming lead come from the interference of the incoming wave 
with the wave that is backscattered from the dot array.
Their absence is a signature of a resonance peak in the transmission spectrum, as there is no overall backscattering and transmission is unity.
It must be noted that the colormap for the DoS in each of the sub-
plots is normalized to the maximal value, such that same colors at different sub-plots
do not represent equal absolute values (which are irrelevant in the present analysis).
Starting with the single oval in the first column [Fig.~\ref{fig4} (a1)-(e1)], 
we see that the incoming wave is reflected at all energies (a) to (e) around $\kappa_p$,
leading to a transmission significantly less than unity ($T^{(1)}(\kappa_p) \approx 0.07$).
When a second oval is added, there is an energy between (a) and (e), namely $\kappa = \kappa_p$ represented by row (c), 
for which the backscattering of the single oval
is cancelled by the presence of the added oval and the connecting bridge:
the wave is multiply reflected between the two ovals through the bridge, 
resulting in a quasi-standing wave along the chain [see (c2)] that constitutes a resonant state for the open system,
leading to a transmission of unity.
A third oval added in front of the two introduces the backscattering again at $\kappa_p$ [see (c3)], 
while the transmitted part from this first oval is perfectly propagated through the remaining two as in (c2).
Thus the transmission in (c3) equals the single dot transmission, $T^{(3)}(\kappa_p) = T^{(1)}(\kappa_p)$.
The backscattering of the third oval at $\kappa_p$ is cancelled by addition of a fourth oval [see (c4)],
just as we went from (c1) to (c2), so that the $T^{(2)}(\kappa_p)$ resonance peak is recovered in $T^{(4)}(\kappa_p)$,
although with a smaller width.
Thus, the sequential cancellation of the backscattered wave leads to the even-odd oscillations of $T^{(N)}(\kappa_p)$
seen in Fig.~\ref{fig3}~(A).
Resonant states are also accessed for 3 and 4 dots in (b3) and (a4) below $\kappa_p$ and in (d3) and (e4) symmetrically above $\kappa_p$.
Similarly, for each number of dots $N$ there are $N-1$ accessible resonant states, including the one at $\kappa_p$ for even $N$,
at energies symmetrically positioned with respect to $\kappa_p$.
Just as the $T^{(2)}$ resonance is recovered in $T^{(4)}$, each $T^{(N)}$ resonance is recovered at multiples of $N$,
where the resonant state in the chain can be decomposed into multiple connected resonant states.
We notice that the two branches of resonances, one below and one above $\kappa_p$, 
are associated with two different leaking eigenstates of the single oval with closed leads - 
they inhabit, for example, the central oval in (b3) and (d3), respectively.
Their interference in the open single oval system forms the scattering wave in column (1).
These three wavepatterns are combined among the $N$ ovals in the open chain, 
to form the $N-1$ resonant states leading to the peaks around $\kappa_p$.
The formation of resonant states occurs similarly around all $T^{(2)}$ resonances of BW type (seen in Fig.~\ref{fig2}).
Characteristically, moving from a $T^{(2)}$ resonance to the next one at higher energy 
adds a node in the quasi-standing wave within the two ovals and the connecting bridge.
Increasing the length of the bridge shifts the resonances to lower energies and reduces the $\kappa$-distance between them, 
as the wavelength in the quasi-standing wave overall increases, 
in accordance to the effective resonator picture described above.

Conclusively, there are two types of resonances to be distinguished in the transmission spectra for the array of $N$ dots:
(\textit{i}) the series of equidistant Fano resonances, arising from the confined single dot excitation modes in the continuum of the channel, 
which are $N$-fold split due to coupling between the ovals, 
and (\textit{ii}) the series of nonequidistant BW resonances, 
resulting from resonant tunneling states that form in the chain, which are $(N-1)$-fold split.

Following the discussion above, we now consider the impact of the perpendicular homogeneous magnetic field on the transport through the device.
When the field is switched on the phases of the different states forming in the ovals are modulated, 
and consequently the interference of the states contributing to transmission changes.
Thus, depending on the field strength, the transmission spectra for the single and multiple dots are accordingly modified.
As we see in Fig.~\ref{fig2}, the weak field of $20~\rm{mT}$ introduces dramatic changes in the spectra.
The slowly varying background of the single oval case is generally raised throughout the channel, 
removing the characteristic suppression around its middle in the absence of the field.
The overall very high transmission is interrupted by series of dips in its lineshape.
The sharp Fano resonances undergo only a very slight energy shift (visible for the Fano resonance in Fig.~\ref{fig3}~(A)~),
because the spatial distribution of the wave function remains practically unaffected by the low field chosen.
Again the multidot chain provides a more complex transmission spectrum, 
resulting from the subsequent matching conditions for the wave function at the connections between the dots.
The BW and Fano resonances are multiply split like in the field free case, 
and dips and plateaus become sharper and more pronounced as dots are added to the chain, saturating into a banded transmission.
In contrast to the field free case, the transmission pattern is now dominated by narrower gaps and wider transmittive bands.
Thus, also for the long chain of dots the overall transmission is drastically raised by the applied field.
A more detailed analysis of the modification of the conductance with varying field will be presented in the next subsection.

\subsection{Conductance switching \label{switch}}
The normalized conductance $g = \frac{h}{2e^2} G$ of the chain of oval dots is shown in Fig.~\ref{fig3}~(B) (dashed curve) for a temperature $\Theta = 0.2~\rm{K}$ 
over a part of the first transmission channel, in direct comparison to the transmission function. 
The parameter $\kappa$ now represents the scaled Fermi energy of the incoming electrons, around which the transmission function is thermally averaged.
At zero temperature conductance and transmission are equal, 
but as $\Theta$ is increased peaks and dips in the spectrum become less pronounced due to the increased range of contributing energies.
As mentioned above, already at the low temperature chosen, the detailed structure of the transmission is essentially lost:
the sharp resonant peaks are washed out, reflecting their negligible contribution to the conductance.
Also the formation of sharp transmittive bands for the multidot chain is relaxed with thermal averaging.
For long interdot leads (Fig.~\ref{fig3}(d),(e)) the conductance features follow the trend of the single dot case, that is, 
it exhibits similar humps in energy, yet with smaller amplitude.
Similar modifications of the transmission spectra through thermal averaging hold for the conductance profile in the presence of the magnetic field.

A key feature of the oval shaped cavity is the formation of the wide suppression valley in the transmission spectrum of the first transversal channel, 
which is essentially retained also for the conductance at low temperature.
In order to demonstrate the suitability of the chain of dots as a magnetically induced conductance switch, 
we exploit the lifting of this suppression when the field is turned on,
aiming at a high ratio of finite- over zero-field conductance.
In the following we optimize the switching ratio taking into account all relevant parameters ($\delta, B, L, N$), 
as well as finite temperature and impurity scattering effects (see subsection III.C).
First we consider the quantity $G_{\rm{off}}^{\rm{min}}$ which is the zero-field finite temperature conductance 
minimized with respect to the position of the Fermi energy in the first channel.

\begin{figure}[b!]
    \begin{center}  
      \includegraphics*[width=0.45\textwidth]{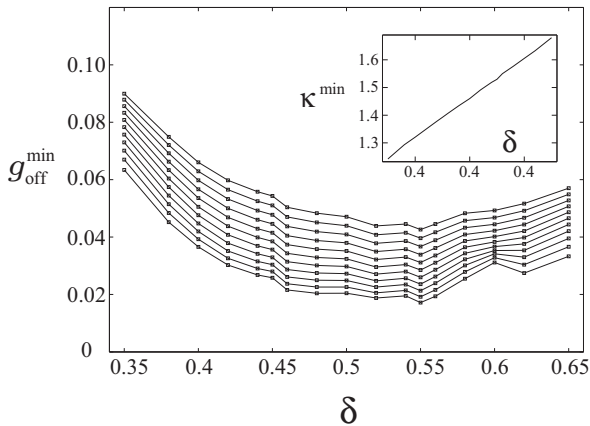} 
     \end{center}
  \caption{Minimal zero-field conductance $g_{\rm{off}}^{\rm{min}}$ (see text) as a function of the deformation parameter $\delta$ for a single dot, at temperatures (bottom to top) $\Theta = 1.0,1.1,...,2.0~\rm{K}$; the inset shows the change of the optimized channel number $\kappa^{\rm{min}}$ with $\delta$ at $\Theta = 2~\rm{K}$ (the dependence is the same for the other temperature values).}
  \label{fig5}
\end{figure}

In Fig.~\ref{fig5}, $g_{\rm{off}}^{\rm{min}} = \frac{h}{2e^2}G_{\rm{off}}^{\rm{min}}$ is plotted as a function of $\delta$ at different temperatures for a single oval dot.
We see that an optimal value for $g_{\rm{off}}^{\rm{min}}$ is obtained around $\delta = 0.5$, with a small dip at $\delta = 0.55$, 
while it increases for larger or smaller deformation of the oval.
It must be noted here that the modification in the spatial extension of the oval for a change $\Delta\delta \approx 0.05$ is of the order of 1\%,
a challenging accuracy for an experimental realization of the device.
We therefore keep the roughly optimized value of $\delta = 0.5$ as a reference for the following analysis.
As shown in the inset of Fig.~\ref{fig5}, the channel number $\kappa^{\rm{min}}$
of this minimum depends approximately linearly on $\delta$ 
where the corresponding Fermi energies are located close to the center of the first channel.
For $\delta = 0.5$ we have $\kappa_c = \kappa^{\rm{min}}(\delta = 0.5) \approx 1.46$ in the single dot case ($N = 1$).
This shift of the optimal Fermi energy, that holds for all temperatures considered, 
is due to the modification of transversal modes inside the dot, 
which are shifted to higher energies as the oval becomes narrower with increasing $\delta$.

\begin{figure}
    \begin{center}  
      \includegraphics*[width=0.48\textwidth]{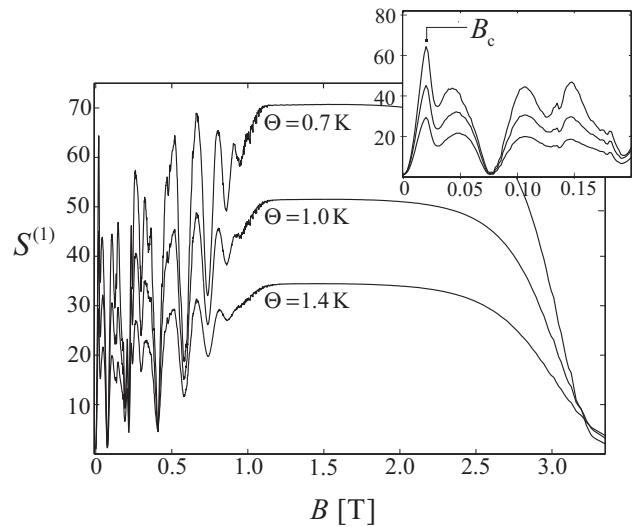} 
     \end{center}
  \caption{Single dot switching ratio $S^{(1)}$ at $\kappa = \kappa_c \approx 1.46$ as a function of the magnetic field, for 		
           temperatures (top to bottom) $\Theta = 0.7, 1.0, 1.4~\rm{K}$;
	   the inset shows the irregular oscillations for low field strengths.}
  \label{fig6} 
\end{figure}

The single dot switching ratio $S^{(N = 1)}(B) = G_{\rm{on}}^{(N = 1)}(B) / G_{\rm{off}}^{(N = 1)}$ at $\kappa = {\kappa_c}$
is shown in Fig.~\ref{fig6} for varying magnetic field strength at different temperatures.
As $S^{(1)}(B)$ equals the finite field conductance normalized to $G_{\rm{off}}^{(1)}$, 
it describes the changes of the conductance induced by the field.
For low field strengths (inset of Fig.~\ref{fig6}) the modulation of the phase of the longitudinal states in the dot leads to Aharonov-Bohm (AB) like oscillations in the conductance.
At the energies we consider here, only three of these leaking states are present \cite{Buchholz2008}.
However, the presence of more than two channels inside the dot gives rise to the superposition of magnetoconductance oscillations,
so that $S^{(1)}(B)$ loses the periodicity expected for AB oscillations of a 1D quantum ring. 
As the field strength is increased, apart from their phase, also the spatial distribution of the states in the dot is affected.
Confined states are eventually deformed into leaking ones, opening further channels for the transmission.
The first magnetoconductance peak at $B_c \approx 0.02~\rm{T}$ is seen to be the highest in the low field regime, 
giving a switching ratio of $S^{(1)}(B=B_c) \approx 65$ at $\Theta = 0.7~\rm{K}$.
For higher field strengths the transmittive states are gradually localized into edge states (with Larmor radius $\lesssim R/4$ for $B \gtrsim 0.8~\rm{T}$) along the border of the cavity, all within the first magnetic Landau level \cite{Rotter2003}.
Following the edges of the billiard, the electrons are now more easily transmitted, resulting in an increased overall conductance.
At a field strength of $B \approx 1.2~\rm{T}$ these modes become perfectly transmittive along the edges of the structure,
and the switching ratio reaches a plateau of maximal value.
For even higher magnetic field strength the transmission decreases drastically
as the incoming electrons gradually fail to overcome the magnetic barrier provided by the first Landau level,
and the conductance drops to zero.
At higher temperatures the features of the magnetoconductance remain; 
however, as a broader energy window with higher transmission parts is contributing to the thermal averaging,
the switching ratio is generally lowered, because $G_{\rm{off}}^{(1)}$ increases.
Also the amplitude of the oscillations decreases with temperature,
as the magnetically induced changes in the detailed structure of the transmission have a smaller impact on average.
For $N > 1$ the magnetoconductance behaves similarly, but the switching ratio overall acquires higher values, 
because of the even lower zero-field conductance, resulting from the formation of gaps in the transmission spectra.

\begin{figure}[ht!]
    \begin{center}  
      \includegraphics*[width=0.45\textwidth]{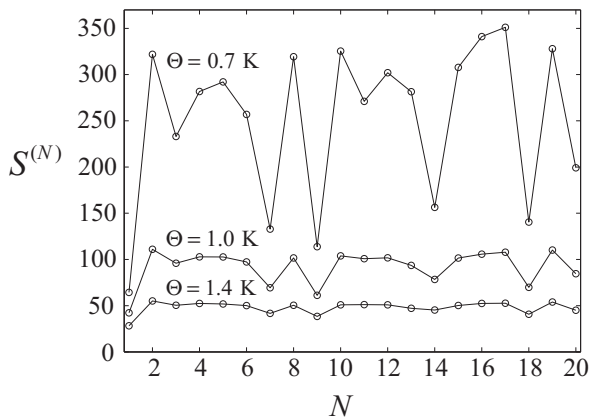}
     \end{center}
  \caption{Switching ratio $S_{c}^{(N)}$ for varying number of dots $N$ with connecting bridge length $L = W$, 
           for different temperatures $\Theta$.}
  \label{fig7} 
\end{figure}

The magnetoconductance is calculated for spinless particles and hence does not describe electronic transport for high magnetic field strengths. 
But, as we are aiming at a high switching ratio, we concentrate in the following on the first maximum  $S_{c}^{(N)} = S^{(N)}(B=B_c)$, 
which occurs approximately at the same field strength $B_c \approx 20~\rm{mT}$
for all considered numbers of dots $N$.
For this weak magnetic field we can neglect the Zeeman splitting.
In Fig.~\ref{fig7}, $S_{c}^{(N)}$ is presented for a varying number of dots in the chain, again at different temperatures.
We allow for the parameter $\kappa_{c}^{(N)}$, which represents the scaled Fermi energy of the incoming electrons, 
to be optimized individually to minimize $G_{\rm{off}}^{(N)}$ for each number $N$.
At sufficiently low temperature, by connecting a second oval to the single one 
we gain a substantial factor with respect to the increase from $S_{c}^{(1)}$ to $S_{c}^{(2)}$ (~$\approx 320$ for $\Theta = 0.7~\rm{K}$),
which, as pointed out, results from the lower zero-field conductance.
For $N > 2$ the switching ratio fluctuates around a temperature dependent mean value, 
due to its high sensitivity with respect to the optimized $G_{\rm{off}}^{(N)}$ at low temperatures, which changes  for each $N$.
At higher temperatures the fluctuations are weakened, but $S_{c}^{(N)}$ is then also lowered drastically.

It is obvious that the optimization of the switching ratio 
strongly depends on the temperature:
High switching ratios require low temperatures, $\Theta \lesssim 2~\rm{K}$ for our setup.
Nevertheless, we see that the current switching functionality of the device is significantly enhanced throughout the temperature range considered,
by taking e.g. two dots instead of a single one.

\subsection{The impact of impurities\label{impurities}}

Let us explore the impact of impurity scattering, i.e. disorder, on the magnetoconductance.
This is implemented in the form of remote impurity scattering in the presence of a modulation-doped layer above the 2D structure.
We consider pointlike negatively charged impurities of 2D density $n_{\rm{imp}}$ distributed on a plane at distance $d$ above the 2D electron gas (2DEG), excluding them from the region of the semi-infinite leads.
The plane is partitioned into small pieces of area $1 / n_{\rm{imp}}$, 
within each of which one impurity is placed at random position, thus constituting a quasi-random distribution of impurities, 
with an upper bound on their local concentration.
The electrostatic potential of each impurity is screened by the 2DEG at the plane of the device structure, 
so that the effective potential that an electron feels at distance $r$ from the impurity is modeled by \cite{Davies1997}
\begin{eqnarray}
V_{\rm{scr}}(r) = \frac{A(d)}{r^3}
\end{eqnarray}
where
\begin{eqnarray}
A(d) = \frac{e^2}{4\pi\epsilon_0\epsilon_b}\frac{q_{\rm{TF}}(1+q_{\rm{TF}}d)}{q_{\rm{TF}}^3}
\end{eqnarray}
with $\epsilon_b$ denoting the relative permittivity of the material.
The Thomas-Fermi screening wave number $q_{\rm{TF}}$ is, for the low temperatures considered, approximated by $q_{\rm{TF}} \approx 2/a_{\rm{B}}$,
where $a_B$ is the effective Bohr radius.
As typical values for a GaAs semiconductor we take $\epsilon_b = 13.8$ and $a_B = 9.8~\rm{nm}$.

As the distance $d$ of the impurity layer is made very short ($d \lesssim 30~\rm{nm}$ in the present scaling), 
the corresponding transmission spectra (not shown here) are drastically changed with respect to the clean case (see Fig.~\ref{fig2}),
as a result of the influence of the impurity potential on the transport through the device.
The randomized potential landscape in the dot chain leads to a spatial deformation of the existing states and
a breaking of the symmetries present in the clean system:
The sharp Fano resonances are shifted due to the perturbation of the confined eigenstates in each dot,
differently for each individual impurity configuration.
The impurity potential also changes the energies of the leaking states,
which results in modified conditions for their coupling to the leads,
so that the broad transmission maxima are shifted, too.
Additionally, new transmission peaks are introduced by leaking states that did not contribute in the clean case due to their symmetry \cite{Buchholz2008}.
For not too short impurity layer distance though, the described suppression valley in the conductance of the clean system is retained, 
still making it sensible to speak about magnetic conductance switching.
The effects of disorder are of course enhanced with increasing impurity density; we use here a value of $n_{\rm{imp}} = 0.0025~{\rm{nm}}^{-2}$. 
This rather high density of remote impurities is employed here in order to intensify their impact on transport in our simulations, whereas in practice cleaner samples are realizable for use in semiconductor nanostructures \cite{Buchholz2009,Friedland1996}.

\begin{figure}
    \begin{center}  
      \includegraphics*[width=0.45\textwidth]{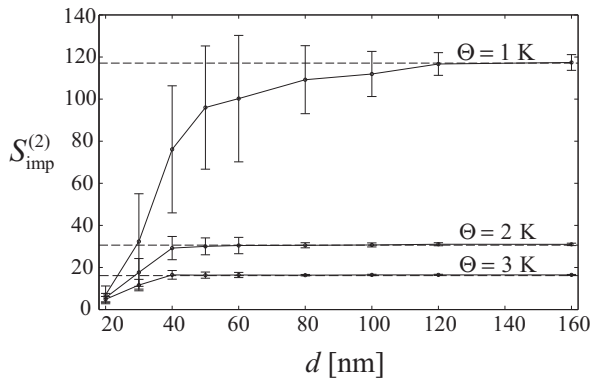} 
     \end{center}
  \caption{Configuration mean and standard deviation of the switching ratio $S_{\rm{imp}}^{(2)}$ of two connected dots  
           as a function of the impurity layer distance $d$ for different temperatures.
           The dashed lines give the values of the disorder free case.}
  \label{fig8}
\end{figure}

In Fig.~\ref{fig8} the switching ratio is shown as a function of the distance $d$ from the impurity layer for two connected ovals.
The values of $S_{\rm{imp}}^{(2)}$ for each $d$ are the average over 27 configurations of the randomly distributed impurities.
When the impurity layer is closer to the 2D conducting structure, the average switching ratio is in general lower than its value in the clean system,
the latter being practically reached for a distance $d \gtrsim d_0(\Theta)$, depending on the temperature.
For $\Theta \approx 1~\rm{K}$ we have $d_0 \approx 100~{\rm{nm}}$, corresponding to a transport mean free path $l_{tr} \approx 24~\rm{\mu m}$ in the first Born approximation \cite{Davies1997}.
Nevertheless, the relatively large deviations from the mean indicate that, for each $d \lesssim d_0(\Theta)$, 
there are certain impurity configurations that provide a switching ratio much higher or lower than the average.
This is due to the high sensitivity of ${G_{\rm{off}}}$ with respect to the potential pattern that is formed on the plane of the array.
If the impurity configuration is, for example, such that a potential maximum is blocking the opening of a cavity to a lead,
then ${G_{\rm{off}}}$ is suppressed, as the wave coming from the lead is strongly backscattered.
This backscattering can be lifted when the magnetic field is turned on, leading to an overall 
increased switching ratio for this configuration.
On the other hand, when the configuration of the impurities does not block the leads, 
${G_{\rm{off}}}$ in the suppression valley is slightly higher compared to the clean case due to the additional resonances in the transmission, causing a reduced switching ratio.
Thus, at distances where the potential on the 2DEG plane is not too strong to permit transmission at all, 
the randomly distributed impurities lie within a broad variation between the cases of blocking and non-blocking configurations,
keeping the deviations from the mean high.
When the impurities are put too close to the 2D structure ($d \lesssim 30~\rm{nm}$), 
the shape specific suppression feature of the zero-field transmission is essentially lost,
so that the overall conductance is practically unaffected by the field strength,
which thus minimizes the switching effect.
For larger impurity layer distances the mean $S_{\rm{imp}}^{(2)}$ eventually saturates into the clean case value with decreasing deviations,
as the potential becomes too weak to affect the transmittive states in the dots.

Using random impurity distributions to investigate the functionality of magnetic current switching in a more realistic environment,
one can speak of a temperature dependent lower bound of the switching ratio (see Fig.~\ref{fig8}) depending on the specific setup.
This lower bound is increased as the influence of disorder is suppressed, that is, when a longer mean free path for the electrons is achieved.
Technological progress actually makes it feasible to reach mean free paths in heterostructures comparable to the size of
realizable nanoscale devices \cite{Reuter2001,Knop2005,Buchholz2009}.
The almost ballistic nature of electron transport then allows for controllable conductance switching at low temperatures, 
in the sense that it is determined by the specific shape of the conducting device, the electron energy and the applied magnetic field.

\section{Conclusions\label{conclusions}}

Having investigated the transmission properties of a linear array of equidistant identical oval shaped quantum dots, 
we demonstrated the functionality of such a structure as a magnetically controlled switching device in the deep quantum regime.
The switching effect arises from the lifting of a deformation specific suppression in the transmission of the oval 
when a weak perpendicular field is turned on.
The suppression valley in the transmission results from the destructive interference of states in the dots 
that are strongly coupled to the leads,
and is specific to the elongated shape of the single billiard.
This makes the effect relevant in systems of similarly shaped dots (e.g. elliptical).
The switching ratio oscillates with the magnetic field strength, 
but as the effect is prominently present even at very weak fields, 
we have concentrated on its first peak.
We have shown that the extension of the single dot into a chain of dots causes a much higher switching ratio, 
due to a stronger suppression of the zero-field conductance.
However, we point out that almost optimal switching can be obtained by connecting only one more dot to the single one,
giving a multiple value for the switching ratio while keeping the system size small.
This could make the device practically advantageous but also favors quantum coherence itself, 
which is the principal requirement for the interference effects to take place.
The efficiency of switching is lowered with increasing temperature, 
as the desired shape specific characteristics of the transmission spectra are thermally washed out,
which poses a limitation to low temperatures (up to about 2 Kelvin).
In spite of the possibility to achieve mean free paths of the 2DEG much longer than the extent of the studied system,
we have additionally investigated the robustness of the switching ratio in the presence of impurity scattering.
The switching ratio acquires a higher or lower value than in the clean case depending, respectively, 
on whether the impurity configuration is blocking transport at zero magnetic field or not.
Thus, for randomly distributed impurities a temperature dependent lower bound for the switching ratio of a sample can be set.
The efficiency of magnetoconductance tuning then remains to be specified for the individual device.
Conclusively, it is demonstrated that electron billiards of specific geometry and chains thereof
can be used, due to regularities in the suppression of their transmission, to design low temperature magnetoconductance. 

This work was supported by the German Research Foundation (DFG)
within the framework of the Excellence Initiative through the
Heidelberg Graduate School of Fundamental Physics (GSC 129/1).
D.B. acknowledges financial support of the DFG within the International 
Research Training Group: Complex Processes: Modeling, Simulation and 
Optimization. (IGK 710)

\bibliographystyle{prsty}
\bibliography{./literature}

\end{document}